\documentstyle[11pt,gh2001-asp,twoside,epsf]{article}
\def\edcomment#1{\iffalse\marginpar{\raggedright\sl#1\/}\else\relax\fi}
\reversemarginpar

\begin{document}
\title{Star formation along the Hubble sequence: results from a new H$\alpha$ Galaxy Survey}
\author{Neville S. Shane and Phil A. James}
\affil{Astrophysics Research Institute, Liverpool John Moores University, Egerton Wharf, Birkenhead,CH41 1LD, UK}

\begin{abstract}
We are conducting a large survey of star formation within $\sim400$
nearby spiral and irregular galaxies through imaging of the H$\alpha$
emission.  We present here some of our first results from 104 of the
galaxies in our sample, investigating the variation of SFR and H$\alpha$
EWs along the Hubble sequence for these galaxies.  We find a strong
dependence of SFR on Hubble type, peaking for Sbc galaxies, but with
large dispersions within each type.  There is a possible dependence of
SFR on bar presence, but none on group membership.  There is an
increase in EWs for later Hubble types, but with large dispersions
within each type.  We find no dependence of EW on bar presence, group
membership or on absolute magnitude.
\end{abstract}

\section{The H$\alpha$ Galaxy Survey}
The H$\alpha$ Galaxy Survey (H$\alpha$GS) is a major survey of star
formation in the local Universe, using the 1m Jacobus Kapteyn
Telescope (JKT) at La Palma, and redshifted H$\alpha$ filters to
determine the quantity and spatial distribution of ionized gas in
$\sim400$ spiral and irregular galaxies within 30 h$^{-1}$ Mpc.  We
are using broad- and intermediate-band R filters for continuum
subtraction.  Our sample was selected, using the Uppsala Galaxy
Catalogue (Nilson 1973) as the parent catalogue, within 5 redshift
shells, but with the angular area of the selection region decreasing
such that all shells have equal volumes.

The well-defined selection criteria, and the large total number of
galaxies, mean that it will be possible to combine the data for all
shells such that galaxies with a wide range of luminosities, diameters
and surface brightness are represented.  Previous studies, e.g. those
of Kennicutt \& Kent (1983) and Young et al. (1996), have been
weighted heavily towards large, luminous and relatively rare Sa-Sc
spirals.
The low luminosity, dwarf galaxies in our sample are the most populous
in the Universe, and our survey will determine whether they are in
fact responsible of a significant fraction of the total star formation
locally.

We have been awarded 14 weeks of observing time over 2000-2002
(long-term status) on the JKT to complete the H$\alpha$ observations.

We are aiming to address such questions as:
\\- What is the total star formation rate (SFR) in the local Universe?  
\\- Which galaxy types and morphologies dominate?  
\\- Does nuclear activity (star formation or AGN) depend on galaxy properties 
and environment?  
\\- What is the star formation distribution within galaxies?
\\- How does environment affect star formation rate?

In this paper we will present preliminary results from SFRs and
equivalent widths (EWs) so far calculated for 104 of the galaxies in
our sample.  These galaxies were all observed under photometric
conditions.

\begin{figure}
\plotfiddle{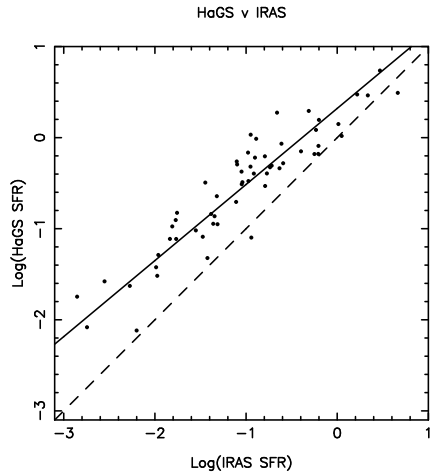}{5.5cm}{270}{33}{33}{-120}{180}
\caption{Comparison of H$\alpha$ to IRAS SFRs}
\end{figure}

\section{Star formation rates}
The total SFR for each galaxy was calculated from the H$\alpha$
luminosity using the conversion of Kennicutt, Tamblyn, \& Congdon
(1994):
\begin{quote}
SFR(M$_{\sun}yr^{-1}$) = L$_{H\alpha}$(W) $\times$ 7.94 $\times$ 10$^{-35}$
\end{quote}
In this preliminary reduction, the luminosities were corrected for
internal extinction assuming a constant value of 1.1mag for each
galaxy (Kennicutt 1983) and the Galactic extinction value given by
Cardelli, Clayton, \& Mathis (1989).  Corrections for contamination by
the [N{\sc ii}] doublet lines which lie either side of the H$\alpha$
line were applied using the H$\alpha$/(H$\alpha$+[N{\sc ii}]) ratios
derived spectrophotometrically by Kennicutt (1983): 0.75$\pm$0.12 for
spirals and 0.93$\pm$0.05 for irregulars.

For the 104 galaxies investigated, the SFRs range between
$\sim0$ and 6M$_{\sun}yr^{-1}$.  75 of these galaxies are classified as
spiral galaxies and the mean SFR for these is 0.525
($\pm$0.100)M$_{\sun}yr^{-1}$.  The mean SFR for the 29 irregular
galaxies is 0.146 ($\pm$0.072)M$_{\sun}yr^{-1}$.

58 of the 104 galaxies also have IRAS far-infrared fluxes.  In Figure
1 we have plotted our H$\alpha$ SFRs against the IRAS SFRs.  The
dashed lines represents an exact match between the two sets of values.
It can be seen that the points do not lie on the line, but
nevertheless form a good correlation.  The solid line is the best fit
through the points and has a slope of 0.83$\pm$0.06.

In Figure 2 we have plotted the mean SFR for each galaxy type.
The error bars show $\sigma/\sqrt{n}$.  We see that there is a strong
dependence of SFR on Hubble type with the highest values being for
Sb-Scd galaxies, peaking at Sbc.  From the error bars, and also from
Figure 3, where we have plotted each galaxy individually, we see that
there is a large variation present in the SFRs amongst galaxies of the
same type.  This variation is most likely real, and not just due to
measurement errors.

In Figure 3 we have split the sample into barred and unbarred
galaxies.  There seems to be a slight trend for late type (Sd-Irr)
barred galaxies to have higher SFRs than late type unbarred galaxies,
however, the full sample of 400 galaxies will be required to
see if this observation is real.

When we split the sample into 71 field galaxies and 33 group galaxies
(as defined by Huchra \& Geller 1982), we find no definite difference
between the star formation properties of the two samples.

\begin{figure}
\plotfiddle{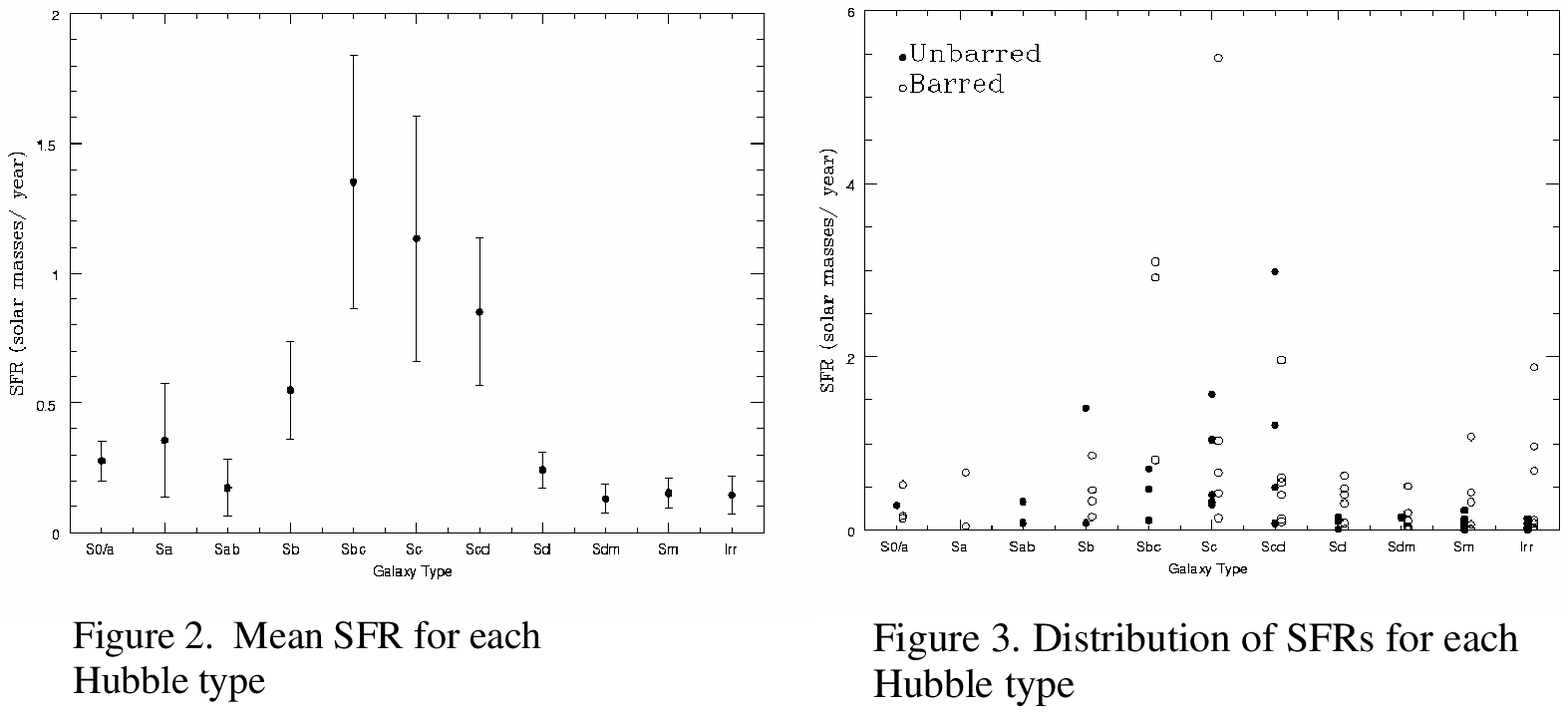}{5.5cm}{0}{84}{84}{-240}{-500}
\end{figure}

\section{Equivalent width measurements}
SFRs are biased by galaxy size.  Larger galaxies are more likely to
have more star forming regions and emit more H$\alpha$.  EW gives a
measure of SFR per unit (red) luminosity, and thus helps to remove
size biases.  EW is also distance independent.

The EWs calculated range between 0.7 and 33nm.  The 75 spirals have a
mean EW of 3.55($\pm$0.30)nm, while the 26 irregulars have a mean value of
5.74($\pm$1.48)nm.  Thus, once we normalise for galaxy size, the irregulars
have similar star formation rates per unit mass to the spirals.

In Figure 4 we have plotted the mean EW for each galaxy type.  We see a
clear trend for later type (Sc - Irr) galaxies to have higher mean EWs than
earlier types (S0/a - Sbc).  From the error bars ($\sigma/\sqrt{n}$) and from
Figure 5, where we have plotted each galaxy seperately, we again see that
there is a large dispersion amongst galaxies of the same type.  These
findings are in agreement with those of Kennicutt (1998).

From Figure 5 we see that there is no detectable dependence on the
presence of a bar (again in agreement with Kennicutt 1998).  We also
find no detectable difference between field and group galaxies.

The two highest EWs, which in fact go off the scale of Figure 5, belong
to UGC3847 (33nm) and UGC3851 (28nm) - an interacting pair of
irregulars.  This finding is in agreement with theories that galaxy
interactions increase the star formation activity of the galaxies
involved.

In Figure 6 we investigate the relationship between EW and absolute V
magnitude.  Gavazzi, Pierini, \& Boselli (1996) find that H$\alpha$ EW
increases steeply with decreasing luminosity.  We find no correlation
overall, although the two highest EW values are found among the low
luminosity irregulars as noted above.

\begin{figure}
\plotfiddle{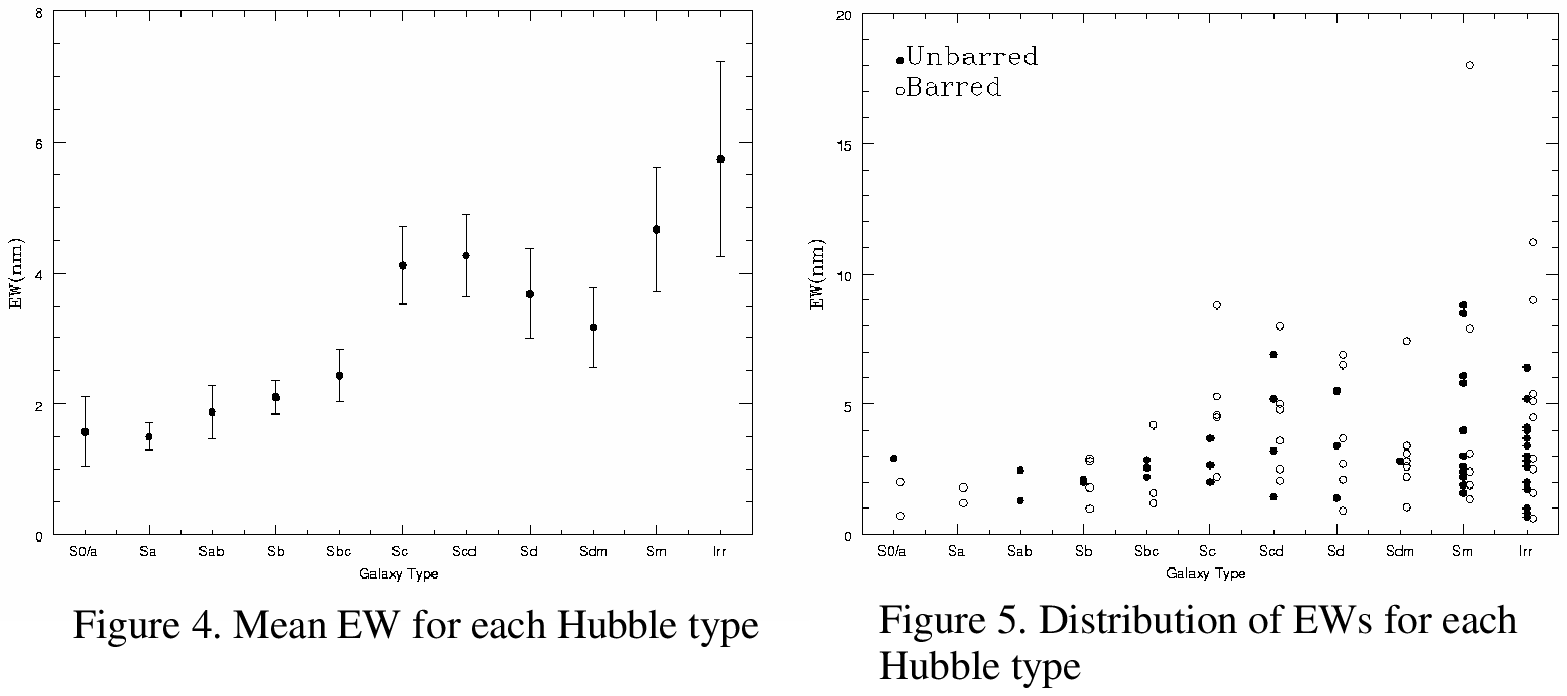}{5.5cm}{0}{84}{84}{-250}{-510}
\end{figure}

\begin{figure}
\plotfiddle{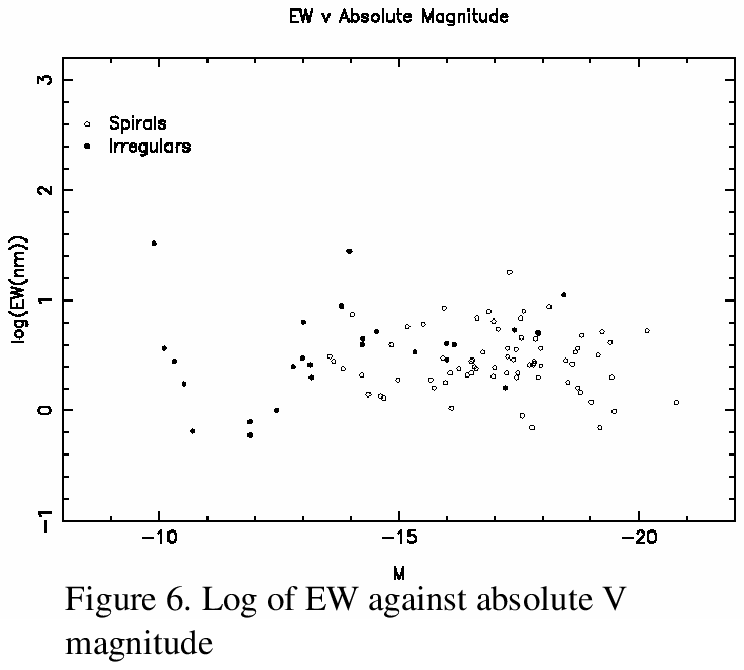}{6.8cm}{0}{100}{100}{-400}{-165}
\end{figure}

\section{Summary}

From a preliminary subset of 104 galaxies ($\sim\onequarter$ of the final sample) we find:
\\- a correlation between our H$\alpha$GS SFRs and IRAS SFRs;
\\- a strong dependence of SFR on Hubble type, but with large dispersions within each type;
\\- a possible dependence of SFR on bar presence for late type galaxies, but none on group membership;
\\- an increase in EWs for later types, but large dispersions within each type;
\\- no dependence of EW on bar presence or group membership;
\\- no correlation between EW and absolute magnitude.

\section{Further work}
The survey is due to finish in January 2002.  With the full sample we
hope to be able to further investigate the above findings as well as
to answer the remaining questions mentioned in \S 1.  

We also aim to derive our own extinction corrections from Br$\gamma$
obervations taken with the WHT and UKIRT, and our own [N{\sc ii}]
corrections using [N{\sc ii}] imaging with the JKT.  In both cases we
will be able to formulate corrections that take into account both
galaxy type and the distribution of dust of [N{\sc ii}] within the
galaxies.  Preliminary investigations show that the [N{\sc ii}]
emission does not trace the H$\alpha$ emission.

\begin{acknowledgements} We would like to acknowledge the input of the members of the H$\alpha$ Galaxy Survey consortium.

This research has made use of the NASA/IPAC Extragalactic Database (NED) 
which is operated by the Jet Propulsion Laboratory, California
Institute of Technology, under contract with the
National Aeronautics and Space Administration.
\end{acknowledgements}

\end{document}